\documentclass{iau}
\usepackage{graphicx}
\usepackage{amsmath}
\usepackage{amssymb}

\title[Uniformity of orientations in astronomy] %% give here short title %%
{A new test of uniformity for object orientations in astronomy}
 
\author[V. Pelgrims]   %% give here short author list %%
{V. Pelgrims}

\affiliation{IFPA, AGO Dept., University of Li{\`e}ge, All{\'e}e du 6 Ao{\^u}t, 17, B4000 Li{\`e}ge, Belgium \\ email: {\tt pelgrims@astro.ulg.ac.be}}

\pubyear{2014}
\volume{306}  %% insert here IAU Symposium No.
\pagerange{119--126}
\jname{Statistical Challenges in 21st Century Cosmology}
\editors{A. F. Heavens, J.-L. Starck \& A. Krone-Martins, eds.}
\begin{document}

\maketitle

\begin{abstract}
We briefly present a new coordinate-invariant statistical test dedicated to the study of the orientations of transverse quantities of non-uniformly distributed sources on the celestial sphere. These quantities can be projected spin-axes or polarization vectors of astronomical sources.

\end{abstract}

\firstsection % if your document starts with a section,
\section{Introduction}

In past decades, large amounts of data have been recorded in various fields of astronomy, propelling poorly known and little understood phenomena into the area of precision science. The cosmic microwave background is one of the most striking example. Such full- and or deep-sky surveys are facing on the problem of huge dataset treatment.
However, aside from these huge observational campaigns, some catalogues contain relevant information for only sparse and non-uniformly scattered sources on the sky.
Appropriate and robust statistical tests have to be dedicated to these small datasets.
To the best of our knowledge, there was a lack of dedicated tests capable of analysing transverse data in astronomy, such as polarizations or projected-spin axes of galaxies or quasars.
Indeed, these quantities are always defined in the plane orthogonal to the line of sight pointing to their corresponding sources through angles relative to one of the basis vectors. Therefore, if we consider sources that are scattered on the celestial sphere, the results of any statistical test performed on these angles will depend on the coordinate system in which the source positions are reported.
When numerous point sources are uniformly scattered the problem is overcome through the use of spin 2 spherical harmonics (in the case of CMB polarization, for instance).
Nevertheless, as soon as sources are rare and non-uniformly distributed over the sky, what is recurrent in astronomy, the problem of tackling the coordinate-system dependence appears.
Forgetting this fact could, in some cases, lead to a misinterpretation of observational data.

In \cite[Pelgrims \& Cudell (2014)]{PC-2014}, we developed a dedicated tool to process this type of sparse data. In the following, we introduce this new statistical test and present an 
application to the sample of optical polarization measurements of 355 quasars non-uniformly scattered on the sky.

\section{The new method}
The basic idea of our method is to consider the quantity which is really measured.
When studying polarization or other transverse data, this quantity is three-dimensional and, the measured axis is defined by the direction of the line of sight $\vec{s}$, where the source is observed, and by the position angle (PA) (defined modulo $\pi$ radians when axial) which indicates the orientation in the plane orthogonal to $\vec{s}$.
In order to study the uniformity of orientations of transverse quantities in a sample, we move from the usual circular data treatment to a spherical data treatment (in the sense of \cite[Fisher (1993)]{Fisher1993} and \cite[Fisher et al. (1993)]{Fisheretal1993}, respectively). Transporting the axes to the origin of the coordinate system, we study their relative orientations through the study of density of points on a unit 2-sphere, where points are the intersections of the axes with the sphere. Simple spherical data analysis, such as those presented in \cite[Fisher et al. (1993)]{Fisheretal1993}, are not applicable in the case of transverse quantities since points are constrained to lie on great circles embedded in planes having the lines of sight as normal vectors. One can nevertheless evaluate the density of points at each location on the unit 2-sphere by adopting Kamb-like methods (e.g. \cite[Vollmer 1995]{Vollmer1995}).

We found convenient to use equal-area spherical caps. Indeed, under the assumption of uniformity of orientations,  this particular shape allows us to predict in a simple way the density of points at each location, evaluating densities through a standard step function.
Namely, the probability ${\ell}^{(i)}$ that the point attached to the source $i$, located in $\vec{s^{(i)}}$, falls inside the cap of half aperture angle $\eta$ and centred in $\vec{c}$ is given by:

\begin{equation}
{\ell}^{(i)}=\begin{cases}
\frac{2}{\pi}\,\mbox{acos}\left(\frac{\cos\eta}{\sin\tau^{(i)}}\right) & \mbox{if}\:\sin
\tau^{(i)}\geq\cos\eta \\
0 & \mbox{otherwise}
\end{cases}\;.
\end{equation}
where $\cos \tau^{(i)} = \vec{c} \cdot \vec{s^{(i)}} $. In fact, adopting a step function, the probability is simply the ratio of the arc length of the great circle embedded inside the cap to that of the great circle itself (i.e. $\pi$, as axial data are under consideration here). The case $\ell^{(i)} = 0$ is well understood through geometrical argument.
It is clear that this probability is independent of the system of coordinates.

Now, considering a sample of sources, one obtains a set of individual probabilities $ \left\lbrace \ell^{(i)} \right\rbrace $ for each location on the sphere. Then, at each location, the Poisson-binomial probability distribution of the density of points is built from the $\ell^{(i)}$'s using a recursive algorithm presented in \cite[Pelgrims \& Cudell (2014)]{PC-2014}, \cite[Chen \& Liu (1997)]{ChenLiu1997} and \cite[Howard (1972)]{Howard1972}.
Then, we are able to test the hypothesis of random orientations at each location of the sphere by evaluating the p-value of the observed density.

For a given sample of sources, the direction of the strongest over-density defines the alignment direction and the corresponding p-value, denoted by $p_{min}$, gives the likelihood of this over-density in this direction.
This value is a \textit{local} probability, it is obtained semi-analytically and it is coordinate-invariant.
A \textit{global} probability is also computed, answering the question of likelihood of such an over-density in arbitrary direction. This probability is evaluated through Monte Carlo simulations by keeping source positions fixed while randomly varying PA according to a uniform distribution.

\section{Application to real data}
\begin{figure}
\begin{center}
\includegraphics[scale=0.24]{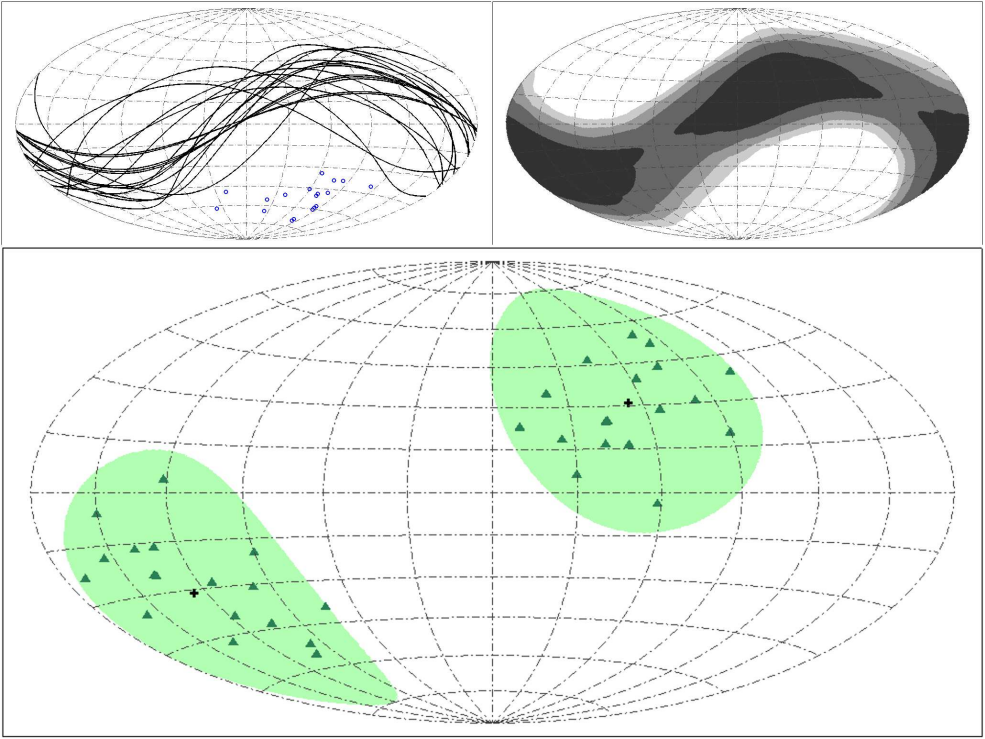}

\caption{The new method at work. Top left: Small circles are source positions while continuous lines are geometrical loci of corresponding polarization points. Note the symmetry as polarization are axial quantities. The probability distributions are computed at each location from the arclength of geometrical loci embedded in the caps. Top right: We build expected density contours from these distributions, taking the mean densities. Dark shades indicate higher density.
Bottom: Observed densities are evaluated by counting the number of polarization points (triangles) falling in each cap. Reporting the observed densities to their corresponding distribution, the alignment direction (crosses) is defined as the centre of the cap (transparent patches) showing the most unexpected over-density.
Hammer-Aitoff projected maps are centred on the Galactic centre with positive Galactic latitude at the top and increasing longitude to the right. The half aperture angle being used is $\eta = 45^\circ$. }
\label{fig:IllustrationMethod}
\end{center}
\end{figure}

In \cite[Pelgrims \& Cudell (2014)]{PC-2014}, we applied this new statistical test to the sample of 355 optical polarization measurements of quasars for which unexpected large-scale correlations have been previously reported in \cite[Hutsem{\'e}kers (1998)]{Hutsemekers1998} and \cite[Hutsem{\'e}kers et al. (2005)]{Hutsemekers2005}.
Performing an analysis on the data sample with our test of density, we confirmed these reported large-scale anomalies and, we proceeded to a precise identification of the quasar regions showing aligned polarization vectors.
Illustration of the new test for one of these regions is shown in Fig.~\ref{fig:IllustrationMethod}. Local and global probabilities for this region of quasars are found to be $1.9 \times 10^{-6}$ and $1.0 \times 10^{-5}$, respectively.

\section{Acknowledgements}
V.P. thanks J.R. Cudell and D. Hutsem{\'e}kers for text enhancement.
This work was supported by the Fonds de la Recherche Scientifique - FNRS under grant 4.4501.05.

\end{document}